\documentclass[showpacs,showkeys,preprint]{revtex4}
\usepackage{amsmath}
\usepackage{amssymb}
\usepackage{amsthm}
\usepackage{graphicx}
\usepackage{dcolumn}
\newcommand{\be}{\begin{equation}}
\newcommand{\ee}{\end{equation}}
\newcommand{\ba}{\begin{eqnarray}}
\newcommand{\ea}{\end{eqnarray}}

\begin{document}
\title{The zero-temperature phase diagram of \\
soft-repulsive particle fluids}
\author{Santi Prestipino$^1$~\cite{aff1}, Franz Saija$^2$~\cite{aff2},
and Gianpietro Malescio$^1$~\cite{aff3}}
\affiliation{
$^1$ Universit\`a degli Studi di Messina, Dipartimento di Fisica,
Contrada Papardo, 98166 Messina, Italy \\
$^2$ CNR-Istituto per i Processi Chimico-Fisici, Contrada Papardo,
98158 Messina, Italy
}
\date{\today}
\begin{abstract}
Effective pair interactions with a soft-repulsive component are
a well-known feature of polymer solutions and colloidal suspensions,
but they also provide a key to interpret the high-pressure behaviour
of simple elements. We have computed the zero-temperature phase diagram
of four different model potentials with various degrees of core softness.
Among the reviewed crystal structures, there are also a number of non-Bravais
lattices, chosen among those observed in real systems. Some of
these crystals are indeed found to be stable for the selected potentials.
We recognize an apparently universal trend for unbounded potentials,
going from high- to low-coordinated crystal phases and back upon
increasing the pressure. Conversely, a bounded repulsion may lead to
intermittent appearance of compact structures with compression
and no eventual settling down in a specific phase.
In both cases, the fluid phase repeatedly reenters at intermediate
pressures, as suggested by a cell-theory treatment of the solids.
These findings are of relevance for soft matter in general, but they
also offer fresh insight into the mechanisms subtended to solid
polymorphism in elemental substances.
\end{abstract}
\maketitle
%
%
\section{Introduction}

Soft-matter systems, like solutions of polymer chains or dispersions of
colloidal particles, 
have been the subject of increasing interest in the last few decades owing
to the possibility they offer of exploring new kinds of equilibrium phase
behaviour which radically depart from the simple-fluid paradigm as
exemplified by rare gases~\cite{Likos1,Malescio1}.
In the reduced-Hamiltonian approach, all of the colloid degrees of freedom
are integrated out, except for the centre-of-mass coordinates
which are taken to interact via an effective potential energy $U$. A further
simplification occurs upon restricting the form of $U$ to a sum of
two-body spherically-symmetric contributions, $U=\sum_{i<j}u(r_{ij})$,
$r_{ij}$ being the distance between particles $i$ and $j$.
It turns out that the $u(r)$ potential is typically softer
(i.e., characterized by a less
steep short-range repulsion) than, say, the Lennard-Jones potential,
if not even being finite for vanishing interparticle distance
as a result of full interpenetrability of the mesoscopic particles.

A well-studied case is that of star polymers, i.e., colloidal particles
with a more or less hard inner core surrounded by a soft penetrable corona
made of grafted polymer chains. For those distances where coronas
overlap, the effective repulsion between two stars grows only
logarithmically with reducing distance, leading to stabilization of
low-coordinated crystals in a range of densities~\cite{Watzlawek,Likos2}.
Another much-studied system is self-avoiding polymers, which are
instead fully penetrable and roughly described by a simple Gaussian
repulsion~\cite{Louis}. At low temperature, the Gaussian-core model
(GCM) can exist in two distinct crystal phases of the face-centred cubic
(fcc) and body-centred cubic (bcc) type~\cite{Stillinger,Lang,Prestipino1}.
Moreover, the fluid phase regains stability on increasing pressure
at constant temperature (reentrant melting) due to an overwhelmingly
larger number, at same energy, of spatially disordered configurations
over crystalline ones.

Simple elements under extreme thermodynamic conditions provide
another instance of a soft short-range repulsion between the system
constituents~\cite{Malescio2}. In this case, the softness of the effective
repulsion is ultimately a reflection, on a coarse-grained level, of
a pressure-dependent atomic radius, as determined in turn by
pressure-induced charge transfer between atomic orbitals.
These effects are well understood in the case of alkali metals,
whose electronic structure at high pressures departs radically
from nearly free-electron behaviour and their common low-pressure
symmetric structure (bcc) becomes unstable to pairing of the ions.
While it could generally be argued that the use of a classical
interatomic potential is not permitted for most elemental liquids
and solids, owing to the fact that electronic effects are important
and are even strongly enhanced by the pressure, a classical framework
for the study of the phase behaviour of simple substances is still
possible, in the same spirit of the Born-Oppenheimer separation
between nuclear and electronic degrees of freedom in molecules.
Non-adiabatic effects are truly negligible for insulators and
semiconductors but seemingly not so for metals where the absence of
a gap between occupied and unoccupied states would make the adiabatic
approximation not particularly well justified.
However, the fraction of electrons that can be scattered by phonons
is rather small even at room temperature since electrons that lie
sufficiently far from the Fermi surface remain frozen in their states
by Pauli's principle. Hence, the notion of an adiabatic potential
as well as the very same concept of a crystal remain meaningful
also for normal metals. This conclusion can be made rigorous by
a theorem due to Migdal~\cite{Migdal}.

In a recent publication~\cite{Malescio2}, we have sketched the overall
phase diagram of a system of point particles interacting through the exp-6
(Buckingham~\cite{Buckingham}) potential, which is being used since long
time as an effective description of rare gases and metals under extreme
conditions~\cite{Ross1,Saija1,Ross2}. In spite of such popularity, important
features of the exp-6 phase behaviour had
previously passed unnoticed. In particular, the reentrant-melting behaviour,
similar to the GCM model, and the rich solid polymorphism, both a generic
trait of elemental substances at high pressures~\cite{McMahon1}.
In the exp-6 system, reentrant melting and solid polymorphism are both
manifestations of the statistical competition, as a function of pressure,
between two distinct scales of nearest-neighbour distance~\cite{Malescio2}.
These two length scales arise as a
result of the radial dependence of the exp-6 repulsion, which shows a
range of distances where the force strength {\it diminishes} with
decreasing interparticle separation (indeed, a rather strong form of core
softness). Whence two preferred values of the mean neighbour distance and the
ensuing frustration of crystalline packing of the standard (fcc or bcc)
type in a certain pressure range.

In this paper, we shall be concerned with the relation between softness of
the interparticle repulsion and polymorphism of the solid phase, with an
emphasis on the occurrence of thermodynamically stable non-Bravais crystals.
We consider four distinct model potentials that have already appeared in the
literature, so as to cover a wide range of possibilities. Aside from the
degree of softness, the chosen potentials differ as for having or not a
hard core as well as for diverging or not at zero separation.
In all cases, we provide an accurate analysis of the zero-temperature
($T=0$) phase diagram by examining a large number of potentially relevant
structures, much larger than ever considered before (at least for these
potentials). In particular, we provide an update of the $T=0$
behaviour of the exp-6 potential as being reported in Ref.\,\cite{Malescio2}.
We know of only a few studies of a similar kind where the search for
stable structures was carried out more systematically (by e.g. a genetic
algorithm)~\cite{Gottwald1,Gottwald2,Fornleitner,Pauschenwein1,Pauschenwein2},
but neither regarded the potentials hereby analysed.

The rest of the paper is organized as follows: In Section 2, we introduce
the potentials for which the $T=0$ phases are computed.
Then, in Section 3, we give a brief description of the crystal structures
being analysed and briefly outline the numerical technique that we use to
optimize a specific structure for a given pressure. Results for the
different potentials are presented in Section 4, with an attempt to find
a common thread among them. Some further remarks and conclusions are
given in Section 5.

%
%
\section{Soft-core potentials}
\renewcommand{\theequation}{2.\arabic{equation}}

Model potentials that describe the effective pair interaction between
particles of simple atomic fluids have a short-range repulsive component
whose steepness (i.e., absolute slope) diverges at zero separation as an
effect of the increasing hindrance, implied by the exclusion principle,
to making particles more and more close to each other. In Lennard-Jones
and inverse-power fluids, the repulsive force ($\equiv -u'(r)$) steadily
increases, with an ever increasing rate, as particles get closer.
By contrast, in soft-core potentials the repulsive force, or at least the
rate with which the force increases, is not a monotonous function of $r$.

The first soft-core potential we want to consider is the exp-6 potential
\be
u_E(r)=\left\{
\begin{array}{ll}
+\infty & ,\,\,\,r<\sigma_M \\
\frac{\epsilon}{\alpha-6}\left\{6\exp\left[\alpha\left(1-\frac{r}{\sigma}
\right)\right]-\alpha\left(\frac{\sigma}{r}\right)^6\right\} &
,\,\,\,r\ge\sigma_M
\end{array}
\right.
\label{2-1}
\ee
where $\epsilon$ is the depth of the attractive well, $\sigma$ is the
position of the well minimum, $\alpha>7$ is a parameter governing the
steepness of the short-range repulsion, and $\sigma_M$ is the point
where the function in Eq.\,(\ref{2-1}) attains its maximum.
This interaction was already studied in some detail for the case
$\alpha=11$ in \cite{Malescio2}, where we also provided a sketch of
the phase diagram at $T=0$.
However, in that paper only a relatively small number of crystal
structures were scrutinized. We here extend that study and find other
crystals with a smaller chemical potential in a certain pressure range.

The exp-6 potential as well as the Gaussian repulsion are instances
of strongly-soft repulsions in that the force slope is {\em positive}
in a range of distances.
However, we can also devise soft-core potentials whose force always
increases upon reducing the interparticle distance while the force steepness
does not monotonously increase as well. This is e.g. the case of the potential
\be
u_{YK}(r)=\epsilon\exp\left\{a\left(1-\frac{r}{\sigma}\right)-
6\left(1-\frac{r}{\sigma}\right)^2\ln\left(\frac{r}{\sigma}\right)\right\}\,,
\label{2-2}
\ee
where $a=2.1$. This potential was first introduced in
Refs.\,\cite{Yoshida1,Kamakura}, and will thus be referred to as the
Yoshida-Kamakura potential. A limited study of the $T=0$ phase diagram
of this model
was already presented in Ref.\,\cite{Yoshida2}. The force $-u_{YK}'(r)$
always increases with reducing $r$ but it does so at a somewhat smaller
rate in a range of distances around $\sigma$ where $u_{YK}''(r)$ develops
a local minimum (see Fig.\,1).

We consider two other model potentials that have recently been
studied in the literature. These are the smoothed hard-core plus
repulsive-step (Fomin) potential~\cite{Fomin},
\be
u_F(r)=\left(\frac{\sigma}{r}\right)^{14}+\frac{\epsilon}{2}
\left[1-\tanh\left(10\,\frac{r-r_0}{\sigma}\right)\right]\,,
\label{2-3}
\ee
with $r_0=1.35\,\sigma$, and the interaction potential of compressed
elastic (Hertzian) spheres~\cite{Pamies},
\be
u_H(r)=\left\{
\begin{array}{ll}
\epsilon\left(1-\frac{r}{\sigma}\right)^{5/2} & ,\,\,\,r\le\sigma \\
0 & ,\,\,\,r>\sigma
\end{array}
\right.
\label{2-4}
\ee
The main reason for considering these potentials is that an accurate
analysis of the phase diagram was performed for both. Reentrant melting
and solid polymorphism are observed in both cases but, as we shall see,
not all stable low-temperature phases were actually identified.
Moreover, while $u_F(r)$ is a strongly-soft potential that, at variance
with the exp-6 potential, lacks a strictly hard core, $u_H(r)$ is a bounded
potential that, unlike the Gaussian repulsion, does not fit our definition
of soft-core potential. This may appear strange since, for bounded
repulsions, two particles can sit on top of each other -- a fact which
by itself is indication of the intrinsic ``softness'' of the interaction.

To highlight the soft nature of the Hertz potential, we look at the
$r$-dependence of what we call the effective inverse-power exponent (EIPE)
of a potential. The EIPE of $u(r)$ is defined as the value of $n$ in
$u_{IP}(r)=A/r^n$ which provides the best local matching between $u_{IP}(r)$
and $u(r)$~\cite{Prestipino2}. By imposing the equality of the functions and
their derivatives at a fixed $r$, we find
\be
n(r)=-\frac{ru'(r)}{u(r)}\,.
\label{2-5}
\ee
In Fig.\,2, we report $n(r)$ for the potentials (\ref{2-1})-(\ref{2-4})
(for the sake of clarity, we have made the discontinuity of $u_E(r)$
at $\sigma_M$ milder by replacing the hard core
with an extremely steep power-law barrier).
We see indeed that, for all potentials, $n(r)$ decreases significantly
for decreasing $r$ in a range of $r$ values, thus signalling core
softening and pointing at the same time to the existence of similarities
among the four models as for the low- to moderate-pressure phase behaviour.

We now prove, by adapting an argument originally due to
Stillinger~\cite{Stillinger}, that a system of particles interacting
through either $u_{YK}(r)$ or $u_F(r)$ is found in a stable fcc crystal
for sufficiently low densities and temperatures,
as indeed suggested by the divergence of the EIPE at infinity.
Let $u(r)\equiv\epsilon\psi(r/\sigma)$ be a purely repulsive and
monotonously decreasing potential. The Boltmann's factor for a pair
of particles is
\be
b(r_{ij})=\exp\left\{-\frac{\psi(r_{ij}/\sigma)}{\psi(R/\sigma)}\right\}\,,
\label{2-6}
\ee
with $R=\sigma\psi^{-1}(k_BT/\epsilon)$, $k_B$ being Boltzmann's constant
(note that the distance $R$ diverges as the temperature $T$ goes to zero).
We now calculate the $T=0$ limit of $b(xR)$,
\be
f(x)\equiv\lim_{T\rightarrow 0}b(xR)=
\exp\left\{-\lim_{R\rightarrow\infty}\frac{\psi(xR/\sigma)}
{\psi(R/\sigma)}\right\}\,.
\label{2-7}
\ee
For
\be
\psi(R/\sigma)\sim\left\{
\begin{array}{ll}
\exp\{-6(R/\sigma)^2\ln(R/\sigma)\} & \,\,\,{\rm
(Yoshida-Kamakura\,\,potential)} \\
\exp\{-\gamma R/\sigma\} & \,\,\,{\rm (Fomin\,\,potential)}\,,
\end{array}
\right.
\label{2-8}
\ee
one invariably has
\be
f(x)=\left\{
\begin{array}{ll}
0^+ & ,\,\,\,0<x<1 \\
1/e & ,\,\,\,x=1 \\
1^- & ,\,\,\,x>1
\end{array}
\right.
\label{2-9}
\ee
In other words, at very low temperature $u(r)$ reduces practically to
the hard-sphere potential and exactly so in the limit $T\rightarrow 0$.
Correspondingly, $R$ plays the role of an effective hard-core diameter.
Since the values of the hard-sphere packing fraction at freezing and
melting are well known~\cite{Frenkel}, the lines of fluid-fcc coexistence
are implicitly given by the equations
\be
\frac{\pi}{6}\left(\rho R(T)^3\right)_f=0.492\,\,\,\,\,\,{\rm and}\,\,\,\,\,\,
\frac{\pi}{6}\left(\rho R(T)^3\right)_m=0.543\,,
\label{2-10}
\ee
which become rigorously valid in the limit $T\rightarrow 0$.

%
%
\section{Crystal lattices and their scrutiny}
\renewcommand{\theequation}{3.\arabic{equation}}

At $T=0$ and fixed pressure, a (crystal) phase is
thermodynamically stable if its chemical potential is smaller than
that of any other phase at equal pressure. We are then faced with
the problem of minimizing the enthalpy per particle as a function of pressure
among {\em all} possible crystals. This is a formidable problem,
since the possibilities are virtually infinite. Hence, we restrict our
search of stable structures to a limited albeit large number of candidates
including, aside from Bravais crystals, also a number of Bravais lattices
with a basis (i.e., non-Bravais lattices) that have been demonstrated to be
relevant for some soft material
or simple substance under high pressures. We do not anyhow consider
the possibility for phases made up of clusters, columns, or lamellae,
which can appear if the
potential exhibits two competing length scales of considerable difference
(see e.g. Ref.\,\cite{Pauschenwein1}). We are fully aware of the
limitations of our approach which, besides assuming a preselected set
of structures, also undergoes increasing difficulties when managing
lattices with many (five or more) characteristic parameters. In this
respect, an approach which employs metadynamics~\cite{Behler,Ishikawa} or a
genetic algorithm~\cite{Oganov,Yao,Abraham} as a tool for an automated search
of optimal solutions inside the structure space
would ensure a higher rate of success.

However, provided the number of crystal lattices being examined is
sufficiently large, direct optimization of structures from a finite set
offers the advantage of weighing up the relative stability between the
optimum solution for a given pressure and the metastable crystals
that crowd around it. Moreover, features like the trend of particle
coordination as a function of pressure or the characteristic distances
at which nearest and next-nearest particles are preferentially located
around a central particle can anyway be grasped by this method.

The crystal lattices that we analyse can be divided in groups of
increasing optimization difficulty, according to the number (from
zero to five) of independent parameters (i.e., axial ratios, angles,
and/or atomic-site parameters), besides the number density $\rho$, that
need to be adjusted in order to minimize the enthalpy at fixed pressure.
For non-Bravais lattices, these parameters are listed in Table 1, along
with the number of inequivalent sites (NIS) of the lattice, defined as
the maximum number of sites whose environments look different as for
the population and/or radius of at least one coordination shell.
For a Bravais lattice one has NIS=1, but there also exist non-Bravais
lattices with this property (e.g. diamond).

The simplest close-packed structures, fcc and hexagonal close packed
(hcp), are well known. In the $[111]$ direction, the stacking sequence
of triangular layers is ABC for fcc and AB for hcp (the bcc lattice can
also be built this way, by suitably spacing the layers one relative to the
other in the same ABC sequence as for fcc). More generally, one can consider
so-called Barlow packings (i.e., the stacking variants of fcc and hcp
packings) where each additional layer involves a binary choice for how to
place it relative to the previous layer. Among them, we consider the
double hcp (dhcp) structure (stacking sequence: ABAC), which is the structure
of $\alpha$La, the triple hcp
(thcp) structure (stacking sequence: ABCACB), and the 9R structure of
$\delta$Sm and $\alpha$Li~\cite{Overhauser} (stacking sequence: ABCBCACAB).
Another simple structure with no parameters
is the diamond lattice (a fcc lattice with
a basis of two atoms), which provides the low-pressure ground-state
configuration of C, Si, Ge, and Sn. Structures with zero parameters
are also the simple-cubic (sc) lattice, the one-species analog of the
fluorite (CaF$_2$) lattice, the A15 lattice (i.e., the structure
of $\beta$W, whose conventional unit cell contains eight atoms), and
the bcc12 lattice (i.e., the structure of the metastable Ga-II phase).

The group of lattices with one free parameter includes, besides a few
Bravais lattices (sh, st, bct, and trig -- i.e., the simple rhombohedral
or hR1 lattice),
also the $\beta$Sn lattice (occurring also for Si, Ge, Rb, and Cs), the
graphite lattice, as well as two cubic lattices obtained by suitably
distorting a bcc supercell (cI16-Li~\cite{Hanfland} and BC8, both with
a conventional unit cell containing eight atoms). hR1 is obtained by
stretching the sc lattice along a body diagonal. The straining parameter
$h$ is usually defined~\cite{CLS} in such a way that $h=0$ gives the sc
lattice, $h=-1/6$ the bcc lattice, and $h=1/3$ the fcc lattice. Both Hg
and Li display a hR1 solid phase. The conventional unit cell of $\beta$Sn
is tetragonal with 4 atoms inside. A cI16 phase has recently been identified
also for Na~\cite{McMahon3,Gregoryanz} while BC8 provides a low-pressure
metastable
phase for Si. The coordinates of the atoms inside the conventional unit cells
of cI16-Li and BC8 are written in terms of a parameter $x$ ranging from 0
(giving back the bcc lattice in the cI16-Li case) to 1/8.

Aside from the Bravais lattices so, sfco, bco, and fco, crystal structures
with two adjustable parameters are also the one-species analog of wurtzite
(ZnS) and the so-called $Imma$ phase of Si (Si-XI, stable between 13 and 16
GPa). The $Imma$ phase is a distortion of $\beta$Sn. Moreover, we consider:
the $\beta$Np structure, which is the one-species analog of PbO (A$_d$, tP4);
the $\alpha$As structure (A7, hR2), also relevant for Sb and Bi, which can
also be viewed as an orthorhombic lattice with $b/a=\sqrt{3}$ and a 12-atom
basis; the $\gamma$Se structure (A8, hP3), also observed in Te; the
$\beta$Mn structure (A13, cP20); and, finally, the C6 structure of Ti, Zr,
and Hf ($\omega$ phase, hP3).

There are two further Bravais lattices with three parameters (sm, sfcm) and
one with five parameters (tric). We also consider another three-parameter
lattice (A20, oC4, $Cmcm$ space group), which provides the structure of
$\alpha$U, $\beta$Ga, and $\gamma$Ti, a four-parameter lattice (A11, oC8,
$Cmca$ space group), and two more five-parameter lattices (oC16-Cs and ST12).
The A11 lattice is the structure of $\alpha$Ga (black P and B above 90 GPa
have a similar structure). A oC8 phase was first predicted for compressed
Li by Neaton and Ashcroft~\cite{Neaton} and later confirmed by Hanfland
{\it et al.}~\cite{Hanfland}. Theoretical calculations by Christensen and
Novikov have predicted the existence of a stable oC8 phase also for Na
above $\approx 250$ GPa~\cite{Christensen}. The orthorhombic oC16 lattice
provides the structure of Cs-V~\cite{Schwarz} and Rb-VI but it is also
observed in Si (between 38 and 45 GPa) and in Ge. Its $Cmca$ structure
contains two types of atom, say A$_1$ and A$_2$, with A$_1$ in planar
arrangements separating A$_2$ double layers. Within the planes, atoms
form a dense packing of dimers. The $Cmca$ structure of oC8 is similar
to oC16 but the double layers of A$_2$ atoms are absent. Of $Cmca$ symmetry
is also the structure of Ca-V~\cite{Fujihisa}, which has the highest
superconducting transition temperature (25 K) among all the elements.
Finally, ST12 denotes a simple-tetragonal metastable phase of Si with
a 12-atom basis.

Let us now briefly outline the procedure we follow in order to optimize
a given crystal structure for assigned potential. Loosely speaking, we
make a grid of points in parameter space on which we search for the
minimum enthalpy at given pressure. We consider lattices with a number
of sites between 8000 and 10000, with periodic boundary conditions.
Lattice sums are extended up to a cutoff distance equal to half the
shortest box length for orthogonal cells (half the shortest distance
of centre to boundary for non-orthogonal cells). Then, a long-range
correction is added to the energy by assuming a radial distribution
function (RDF) of 1 for the distant sites. We start by performing a series
of harsh minimizations on rather coarse meshes but then we refine the
calculation by progressively reducing the mesh size until the lattice
providing the
minimum chemical potential is extracted out of a number of deeper valleys
in parameter space (steepest-descent methods would not necessarily be of
help in this case since one could easily get stuck in local minima). We
feel satisfied when we obtain the density with three exact decimal digits
and the other parameters with two to three decimals.
Occasionally, we resolve near degeneracies by going to larger lattices
and better parameter refining. When more than one crystal structure is
found to give practically the same minimum chemical potential, we check
the identity of the subtended lattices by looking at the discrete RDF of
each.

%
%
\section{Results and discussion}
\renewcommand{\theequation}{4.\arabic{equation}}

The $T=0$ phase diagram of model systems interacting through the
soft-core potentials (\ref{2-1})-(\ref{2-4}) is reported in Tables 2
to 5. Overall, we see the existence of a rich solid polymorphism, with
many exotic lattices providing stable phases at moderately high pressures.
Obviously, we cannot exclude that there might exist other phases which
overcome in stability some of those found (to be sure, one should perform
a molecular-dynamics simulation of the system at fixed pressure, starting
from the high-temperature fluid and cooling it down very slowly until the
solid nucleates -- but the conditions for observing ideal freezing are
hardly achieved in a reasonable amount of computer time).
In this respect, the case of Na
is particularly illuminating~\cite{Gregoryanz}: if the experiment does not
lie, no theoretical study could ever have anticipated the bunch of crystal
phases, some of which extremely complex, that Na exhibits for pressures
above 110 GPa. Yet, we believe that our study gives some general teachings
on the mechanisms underlying spatial ordering at high pressures which would
remain valid even in case some of the phases that we find stable are
actually metastable.

As a preliminary test, we study the Lennard-Jones fluid at $T=0$ and confirm
that it can only exist in two forms, either as a hcp or as a fcc crystal,
the former being stable at lower pressures~\cite{Jackson}. The transition
between the two occurs approximately at $P\sigma^3/\epsilon=800$,
though the chemical-potential difference is very small ($<0.001\,\epsilon$)
for all pressures up to 1000. Moreover, we find that apparently all Barlow
packings are more stable than fcc for $P\sigma^3/\epsilon<800$ while the
metastable phase that is closest in stability to fcc is cI16 (with $x=0.024$
at $P=0$, slightly diminishing on increasing pressure), as suggested by
the simulation~\cite{Eshet}. We also verified that the only $T=0$ phases
of the GCM are fcc and bcc.

Table 2 collects the $T=0$ phases of the exp-6 model with $\alpha=11$.
In Fig.\,3, the chemical potential $\mu$ of the same phases
is plotted as a function of pressure $P$, chosen fcc as reference
(the units of length and energy are set to $\sigma$ and $\epsilon$,
respectively). We do not show $\mu$ for all the scrutinized lattices
since the difference in chemical potential between the stable and
the metastable phases is often very small on the scale of the picture,
insomuch that it would have been difficult to keep track of all
the curves. Examples of a strong competition between different phases
are seen anyway in Fig.\,3, near $P=15000$ (hR1 vs. sh) and near $P=35000$
(the challenge now being among oC8, hR1, sh, and wurtzite).

It transpires from Table 2 that the coordination number $z$ has a regular
trend with pressure. Starting from 12 (fcc-I) at low pressure, $z$ reduces
upon compression down to a minimum of 2 (hR1). Then, it increases
progressively with pressure until becoming, eventually, 12 again (fcc-II).
This behaviour can be rationalized as follows. We see from Table 2 that
the nearest-neighbour distance is $\sigma_M$ for all stable phases except
for fcc-I and bcc-I. Whence the convenience, in order to minimize the energy
at not too low pressure, that the number of first neighbours be as small
as possible. Eventually, however, the $PV$ term in the enthalpy takes over
the energy and there will again be room for close-packed lattices.
The reason why low-coordinated phases do not show up for
systems with a Lennard-Jones-like potential is that, in these systems,
the fcc and bcc crystals anyway manage to accomodate second and third
neighbours at convenient distances. On the contrary, in systems interacting
through soft-core potentials, fcc and bcc local orders are destabilized
by the peculiar dependence of the interatomic force with distance, which,
for high softness degrees, leads to the existence of two incommensurate
length scales that heavily frustrate the too compact arrangements.

Looking at Fig.\,3, a few other comments are in order:
1) the BC8 phase, which was previously~\cite{Malescio2} found to be stable
for pressures between roughly 20000 and 30000, is actually less stable than
$\beta$Sn. 2) Interestingly, there are two
distinct ranges of pressure where the exp-6 system exhibits oC8 order
at $T=0$. In fact, within one of these intervals there are phase transitions
between different oC8 phases, as signalled by the jumps of parameters from
one valley to another in parameter space. We tabulated up to four distinct
oC8 phases for the exp-6 system though, actually, a careful examination of
the data shows that the isostructural transitions, including the weaker ones,
are many more. 3) In a wide pressure interval, the orthorhombic A20 lattice
gives the most stable phase. There are actually five distinct A20 states at
$T=0$ with abrupt transitions between them at specific pressures. These
transitions are in the form of unequal compressions along the $b$ and $c$
axes, also accompanied by an adjustment of the internal parameter $y$.
4) The change in slope which is manifest in all curves at $P\simeq 52000$
is due to a sudden change of the nearest-neighbour distance of the optimal
fcc crystal from roughly $0.406$ down to $0.37381=\sigma_M$
(a similar jump occurs for bcc at the same pressure).
This effect is rather specific to the exp-6 system and ultimately related
to the abrupt change of the potential profile at $\sigma_M$.

Table 3 reports the $T=0$ phase diagram of the Yoshida-Kamakura potential.
A comparison with the exp-6 case reveals a number of
similarities as for the trend of coordination number with pressure
and for the order of appearance of the phases.
Hence, some of the considerations made for the exp-6 model
also apply to this potential, in spite of the fact that the steepness of the
Yoshida-Kamakura potential is a monotonous function of $r$ (see Fig.\,1).
In Fig.\,4, we show the discrete RDF of the $T=0$ phases for $u_{YK}$.
We see that, with the exception of the highest-pressure phases, the range
of distances from a central particle that corresponds to the force
``plateau'' is void of
neighbours, like as if particles tended to sit at the shortest distances
available at same force strength. Something similar occurs for the exp-6
model, see Table 2, where no particles are found in the region of distances
where the potential is concave.

We now try to obtain a rough phase diagram for the Yoshida-Kamakura model
from just the knowledge of its $T=0$ sector. The simplest way
to do this is to lay down a (Lennard-Jones-Devonshire) cell
theory~\cite{LJD} for the solid phases and to use the Lindemann
criterion~\cite{Lindemann} for locating the melting transition.
In the cell theory, a crystal partition function of effectively
independent particles is written down where any given particle, which
can be found anywhere in its own Wigner-Seitz cell (WSC), is acted
upon by the force exerted by the other $N-1$ particles, placed at
their equilibrium lattice positions. In practice,
\be
Z=\frac{1}{\Lambda^{3N}}\int_{{\rm WSC}_1}{\rm d}^3r_1\cdots\int_{{\rm WSC}_N}
{\rm d}^3r_N\,\exp\left\{-\sum_i\phi({\bf r}_i)/(k_BT)\right\}\,,
\label{4-1}
\ee
where
\be
\phi({\bf r})=\frac{1}{2}\sum_{j\neq 1}u(|{\bf R}_1-{\bf R}_j|)+
\sum_{j\neq 1}\left[u(|{\bf R}_1+{\bf r}-{\bf R}_j|)-
u(|{\bf R}_1-{\bf R}_j|)\right]
\label{4-2}
\ee
(we denote by capital letters the positions of lattice sites in the perfect
crystal). Within this theory, the mean square displacement of a particle
is given by
\be
\left<\Delta r^2\right>=
\frac{\int_{\rm WSC}{\rm d}^3r\,r^2\exp\{-(\Phi({\bf r})-\Phi(0))/(k_BT)\}}
{\int_{\rm WSC}{\rm d}^3r\,\exp\{-(\Phi({\bf r})-\Phi(0))/(k_BT)\}}\,,
\label{4-3}
\ee
for $\Phi({\bf r})=\sum_{j\neq 1}u(|{\bf R}_1+{\bf r}-{\bf R}_j|)$.
If we denote $r_{NN}$ the nearest-neighbour distance, the Lindemann fraction
is defined as
\be
{\cal L}=\frac{\sqrt{\left<\Delta r^2\right>}}{r_{NN}}\,,
\label{4-4}
\ee
which is an increasing function of temperature.
The Lindemann rule then states that melting occurs at the temperature $T_m$
where ${\cal L}$ reaches a critical value ${\cal L}_c$ specific of the given
lattice. Experience with other models shows that ${\cal L}_c=0.15\div 0.16$
for fcc and $0.18\div 0.19$ for bcc~\cite{Saija2} while, to our knowledge
at least, there is no general consensus in the literature as to the value
of ${\cal L}_c$ for other lattices (we tentatively assume ${\cal L}_c=0.19$).

Fig.\,5 shows the phase diagram of the Yoshida-Kamakura potential as mapped
out in the way just explained. Aside from the simplicity of the theory used,
in plotting this figure we make two further assumptions: 1) We discard the
possibility that some other phases, not stable at $T=0$, might be promoted
entropically for $T>0$; 2) we exclude that a given phase might invade, for
$T>0$, the density intervals where the adjacent phases are stable for $T=0$.
Looking at Fig.\,5, we see at least one region of reentrant melting between
bcc-I and sh-I, but probably there are others. This is a curious finding in
view of the absence of two clearly-defined repulsive length scales for this
potential (by contrast, reentrant-fluid behaviour is well-documented for
the exp-6 model~\cite{Malescio2} and the Fomin potential~\cite{Fomin}, both
strongly-soft potentials).
Since we do not calculate the phonon spectrum of the solids, we cannot
exclude that some of them are actually mechanically unstable for $T>0$.
However, indirect clue to mechanical stability of a crystal phase can be
the positive value of the elastic constant $K$ associated with the
one-particle potential $\Phi({\bf r})$. This quantity can be extracted
from the ${\cal O}(r^2)$ term in $\Phi({\bf r})$, via a spherical
average~\cite{Yoshida2}.
We have checked that, at least within the range of stability of each phase
in Table 3, the value of $K$ is always positive. Moreover, if we estimate
the mean square displacement of a particle from the harmonic approximation
for $\Phi({\bf r})$, the resulting phase diagram comes out not too
different from that of Fig.\,5.

The $T=0$ phase diagram of the Fomin potential is outlined in Table 4.
We see the same trend of $z$ with pressure as observed in the previous
cases. However, we find only one non-Bravais phase in this case ($\beta$Sn)
which turns out to be almost degenerate with the bct phase throughout the
whole pressure range from 3.5 to 5.5. All in all, the phase behaviour of
unbounded soft-core potentials has some recurrent features
(e.g. rich solid polymorphism, low-coordinated non-Bravais crystal
phases, and reentrant melting) which are also found in the phase
diagram of many simple elements under extreme conditions.
This suggests that the effective two-body (adiabatic) potential of
these substances is a soft-core potential.

Finally, we analyse the $T=0$ phase diagram of the Hertz potential,
Eq.\,(\ref{2-4}), as summarized in Table 5. A rather complex behaviour
shows up for this bounded potential, much more complex than reported in
\cite{Pamies}, especially if compared with the simple phase portrait
of the GCM. In fact, the phases listed in Table 5 are only those stable
for $P<400$ since apparently the sequence of $T=0$ phase transitions
never comes to an end (as observed in Ref.\,\cite{Pamies}, there is
no room for clustering in the Hertz model).
This unique behaviour, never documented before, is the effect of a complex
interplay between energy and volume considerations in the minimization
of enthalpy as a function of pressure. Probably, this behaviour is related
to the absence of a force plateau at $r=0$ which obliges the system to
continuously setting right the positions of the neighbouring particles.

%
%
\section{Conclusions}

In recent years, increasing attention has been devoted to soft-matter
systems as examples of anomalous thermodynamic behaviour, both in and
out of equilibrium. This field of research is very much alive, with many
points of contact with high-pressure physics~\cite{Malescio2}. We have
been focused here on the anomalously rich solid
polymorphism of systems of softly-repulsive particles, by studying the
zero-temperature phase diagram of a number of model pair potentials with
various forms of core softening. We found elements of complexity
that are simply unknown to ``normal'' fluids interacting through a
Lennard-Jones type of potential, with many low-coordinated non-Bravais
lattices providing the structure of stable phases at intermediate pressures.
In a near future, with the advent of new techniques to functionalize the
colloidal surface, one can expect to obtain colloidal systems whose
soft-core potential yields spontaneous assembling into similar exotic
lattices. The relation of solid polymorphism to other kinds of thermodynamic
oddities, such as reentrant melting and water-like anomalies, is the subject
of work in progress.

%
%
\begin{center}
{\bf Acknowledgements}
\end{center}
We acknowledge useful discussions with Ezio Bruno and Sandro Scandolo.
We also thank an anonymous Referee of our previous paper \cite{Malescio2}
for driving our attention to the papers \cite{Yoshida1,Kamakura,Yoshida2}.

\newpage
%
%
\begin{table}
\caption{\label{tab1}
The non-Bravais lattices that have been considered in our search for
stable phases at zero temperature.
For each structure (column 1), the parameter(s) which are needed for a
complete specification of the lattice are reported in column 2. The
symbols in this column have the same meaning as in the web site of
Ref.\,\cite{CLS}, except for $Imma$~\cite{McMahon2} and oC16-Cs~\cite{Schwarz}.
The number of inequivalent sites (NIS, see main text) of the given lattice
is also indicated in column 3.
For Bravais lattices, the parameters are axial ratios ($b/a,c/a$) and
angles ($\alpha,\beta,\gamma$), as usual.
}
\begin{tabular*}{\columnwidth}[c]{@{\extracolsep{\fill}}|c|c|c|}
\hline
lattice & parameters & NIS (with weights, if different) \\
\hline\hline
hcp & --- & 1 \\
\hline
dhcp & --- & 2 \\
\hline
thcp & --- & 2 ($1\times 1,2\times 2$) \\
\hline
9R & --- & 2 ($1\times 1,2\times 2$) \\
\hline
diamond & --- & 1 \\
\hline
fluorite & --- & 2 ($1\times 1,2\times 2$) \\
\hline
A15 & --- & 2 ($1\times 1,2\times 3$) \\
\hline
bcc12 & --- & 1 \\
\hline
$\beta$Sn & $c/a$ & 1 \\
\hline
cI16-Li & $x$ & 1 \\
\hline
BC8 & $x$ & 1 \\
\hline
graphite & $c/a$ & 2 \\
\hline
wurtzite & $(c/a)/\sqrt{8/3},u/(3/8)$ & 1 \\
\hline
$Imma$ & $c/a,\Delta$ & 3 ($1\times 1,2\times 1,3\times 2$) \\
\hline
B10 & $c/a,z$ & 1 \\
\hline
A7 & $b/a,u$ & 1 \\
\hline
A8 & $c/a,x$ & 1 \\
\hline
A13 & $x_1,x_2$ & 2 ($1\times 2,2\times 3$) \\
\hline
C6 & $c/a,z$ & 2 ($1\times 1,2\times 2$) \\
\hline
A20 & $b/a,c/a,y$ & 1 \\
\hline
oC8-Ga & $(b/a)/1.69479,c/a,u/0.1549,v/0.0810$ & 1 \\
\hline
oC16-Cs & $(b/a)/0.594,(c/a)/0.590,x/0.2118,y/0.1781,z/0.328$ & 2 \\
\hline
ST12 & $c/a,x_1/0.0912,x_2/0.1730,y_2/0.3784,z_2/0.2486$ &
2 ($1\times 1,2\times 2$) \\
\hline
\end{tabular*}
\end{table}

%
%
\begin{table}
\caption{\label{tab2}
Exp-6 potential for $\alpha=11$: Zero-temperature phase diagram.
For each phase (column 2), we show the pressure interval of stability
(column 1), the related density interval (column 3), the first- and
second-neighbour distances (column 4), the number of first- and
second-neighbours (column 5), and the values of structure parameters
(column 6).
}
\begin{tabular*}{\columnwidth}[c]{@{\extracolsep{\fill}}|c|c|c|c|c|c|}
\hline
$P_1$--$P_2$ ($10^3\epsilon\sigma^{-3}$) & phase & $\rho_1$--$\rho_2$
($\sigma^{-3}$) & \parbox{0.08\textwidth}{\vspace{2mm}$r_1$ ($\sigma$) $r_2$
($\sigma$)\vspace{1mm}} &
\parbox{0.03\textwidth}{\vspace{2mm}$z_1$ $z_2$\vspace{1mm}} & parameters \\
\hline\hline
0--5.5 & fcc-I & 1.627--6.943 &
\parbox{0.18\textwidth}{\vspace{2mm}0.58838--0.95435 0.83209--1.34966\vspace{1mm}} &
\parbox{0.03\textwidth}{\vspace{2mm}12 6\vspace{1mm}} & --- \\
\hline
5.6--11.8 & bcc & 6.999--9.267 &
\parbox{0.18\textwidth}{\vspace{2mm}0.51947--0.57042 0.59983--0.65867\vspace{1mm}} &
\parbox{0.03\textwidth}{\vspace{2mm}8 6\vspace{1mm}} & --- \\
\hline 
11.9--17.1 & hR1 & 10.123--11.500 &
\parbox{0.18\textwidth}{\vspace{2mm}0.37381 0.53305--0.56628\vspace{1mm}} &
\parbox{0.03\textwidth}{\vspace{2mm}2 6\vspace{1mm}} & $-0.227\div -0.220$ \\
\hline 
17.2--20.4 & oC8-I & 11.982--12.676 &
\parbox{0.18\textwidth}{\vspace{2mm}0.37381 0.52352--0.54667\vspace{1mm}} &
\parbox{0.05\textwidth}{\vspace{2mm}$\simeq 3$ 2\vspace{1mm}} &
\parbox{0.25\textwidth}{\vspace{2mm}$0.85\div 0.91$, $1.24\div 1.38$,
$0.61\div 0.63$, $2.06\div 2.24$\vspace{1mm}} \\
\hline 
20.5--33.3 & $\beta$Sn & 13.455--15.445 &
\parbox{0.18\textwidth}{\vspace{2mm}0.37381 0.52989--0.57802\vspace{1mm}} &
\parbox{0.06\textwidth}{\vspace{2mm}4 $4\div 12$\vspace{1mm}} &
$0.758\div 0.986$ \\
\hline 
33.4--36.9 & oC8-II & 16.464--16.722 &
\parbox{0.18\textwidth}{\vspace{2mm}0.37381 0.50274--0.50809\vspace{1mm}} &
\parbox{0.05\textwidth}{\vspace{2mm}$\simeq 5$ 2\vspace{1mm}} &
\parbox{0.25\textwidth}{\vspace{2mm}$1.10\div 1.15$, $1.81\div 1.85$,
$1.02\div 1.04$, $1.30\div 1.34$\vspace{1mm}} \\
\hline 
37.0--45.3 & oC8-III & 18.561--18.651 &
\parbox{0.18\textwidth}{\vspace{2mm}0.37381 0.54346--0.55677\vspace{1mm}} &
\parbox{0.05\textwidth}{\vspace{2mm}$\simeq 7$ 2\vspace{1mm}} &
\parbox{0.25\textwidth}{\vspace{2mm}$0.95\div 0.98$, $1.08\div 1.09$,
$1.10\div 1.12$, $1.17\div 1.23$\vspace{1mm}} \\
\hline 
45.4--47.6 & oC8-IV & 19.097 &
\parbox{0.18\textwidth}{\vspace{2mm}0.37381, 0.51223\vspace{1mm}} &
\parbox{0.08\textwidth}{\vspace{2mm}$\simeq 7$, 2\vspace{1mm}} &
\parbox{0.22\textwidth}{\vspace{2mm}0.88, 1.11, 1.17, 1.33\vspace{1mm}} \\
\hline 
47.7--48.2 & A20-I & 19.642 &
\parbox{0.18\textwidth}{\vspace{2mm}0.37381, 0.56016\vspace{1mm}} &
\parbox{0.08\textwidth}{\vspace{2mm}$\simeq 8$, 4\vspace{1mm}} &
\parbox{0.25\textwidth}{\vspace{2mm}1.726, 0.666, 0.167\vspace{1mm}} \\
\hline 
48.3--52.1 & A20-II & 19.905 &
\parbox{0.18\textwidth}{\vspace{2mm}0.37381, 0.54115\vspace{1mm}} &
\parbox{0.08\textwidth}{\vspace{2mm}$\simeq 8$, 4\vspace{1mm}} &
\parbox{0.25\textwidth}{\vspace{2mm}1.522, 0.629, 0.179\vspace{1mm}} \\
\hline 
52.2--52.8 & A20-III & 20.074 &
\parbox{0.18\textwidth}{\vspace{2mm}0.37381, 0.53498\vspace{1mm}} &
\parbox{0.08\textwidth}{\vspace{2mm}$\simeq 8$, 4\vspace{1mm}} &
\parbox{0.25\textwidth}{\vspace{2mm}1.468, 0.621, 0.183\vspace{1mm}} \\
\hline 
52.9--57.3 & A20-IV & 20.383 &
\parbox{0.18\textwidth}{\vspace{2mm}0.37381, 0.52665\vspace{1mm}} &
\parbox{0.08\textwidth}{\vspace{2mm}$\simeq 8$, 4\vspace{1mm}} &
\parbox{0.25\textwidth}{\vspace{2mm}1.398, 0.610, 0.189\vspace{1mm}} \\
\hline 
57.4--60.9 & A20-V & 20.813 &
\parbox{0.18\textwidth}{\vspace{2mm}0.37381, 0.51700\vspace{1mm}} &
\parbox{0.08\textwidth}{\vspace{2mm}$\simeq 8$, 4\vspace{1mm}} &
\parbox{0.25\textwidth}{\vspace{2mm}1.327, 0.601, 0.196\vspace{1mm}} \\
\hline 
61.0--68.3 & sh & 22.107 &
\parbox{0.18\textwidth}{\vspace{2mm}0.37381, 0.52864\vspace{1mm}} &
\parbox{0.08\textwidth}{\vspace{2mm}8, 12\vspace{1mm}} & 1 \\
\hline 
68.4--76.9 & wurtzite & 24.337 &
\parbox{0.18\textwidth}{\vspace{2mm}0.37381, 0.52864\vspace{1mm}} &
\parbox{0.08\textwidth}{\vspace{2mm}10, 9\vspace{1mm}} &
$2.225$, $0.734$ \\
\hline 
77.0-- & fcc-II & 27.075 &
\parbox{0.18\textwidth}{\vspace{2mm}0.37381, 0.52864\vspace{1mm}}&
\parbox{0.08\textwidth}{\vspace{2mm}12, 6\vspace{1mm}} & --- \\
\hline
\end{tabular*}
\end{table}

%
%
\begin{table}
\caption{\label{tab3}
Yoshida-Kamakura potential for $a=2.1$: Zero-temperature phase diagram.
Notations are the same as in Table 2.
}
\begin{tabular*}{\columnwidth}[c]{@{\extracolsep{\fill}}|c|c|c|c|c|c|}
\hline
$P_1$--$P_2$ ($\epsilon\sigma^{-3}$) & phase & $\rho_1$--$\rho_2$
($\sigma^{-3}$) & $r_1$ ($\sigma$) & $z$ & parameters \\
\hline\hline
0--0.63 & fcc-I & 0--0.325 & 1.63259-- & 12 & --- \\
\hline
0.64--1.26 & bcc-I & 0.331--0.415 & 1.46282--1.57736 & 8 & --- \\
\hline
1.27--2.29 & sh-I & 0.486--0.594 & 0.90256--0.99613 & 2 & $0.615\div 0.645$ \\
\hline
2.30--2.55 & A7 & 0.614--0.637 & 0.95415--0.96971 & 3 &
\parbox{0.30\textwidth}{\vspace{2mm}$3.962\div 3.975$, $0.181\div 0.184$
\vspace{1mm}} \\
\hline
2.56--4.91 & diamond & 0.670--0.845 & 0.91603--0.98970 & 4 & --- \\
\hline
4.92--5.46 & sh-II & 0.908--0.944 & 0.95874--0.97289 & 6 & $1.381\div 1.388$ \\
\hline
5.47--12.07 & A20 & 0.985--1.322 & 0.90053--0.97273 &
\parbox{0.05\textwidth}{\vspace{2mm}$\simeq 8$\vspace{1mm}} &
\parbox{0.30\textwidth}{\vspace{2mm}$1.728\div 1.731$, $0.626\div 0.646$,
0.167\vspace{1mm}}\\
\hline 
12.08--15.68 & hcp-I & 1.424--1.583 & 0.96311--0.99770 & 12 & --- \\
\hline
15.69--52.75 & bcc-II & 1.593--2.618 & 0.79168--0.93426 & 8 & --- \\
\hline
52.76--138.28 & fcc-II & 2.629--4.024 & 0.70570--0.81328 & 12 & --- \\
\hline
138.29--365.65 & hcp-II & 4.026--6.075 & 0.61516--0.70558 & 12 & --- \\
\hline
365.66-- & fcc-III & 6.077-- & --0.61509 & 12 & --- \\
\hline
\end{tabular*}
\end{table}

%
%
\begin{table}
\caption{\label{tab4}
Fomin potential for $r_0=1.35\,\sigma$: Zero-temperature phase
diagram. Phases with 5 parameters were not included in our scrutiny.
Notations are the same as in Table 2.
}
\begin{tabular*}{\columnwidth}[c]{@{\extracolsep{\fill}}|c|c|c|c|c|c|}
\hline
$P_1$--$P_2$ ($\epsilon\sigma^{-3}$) & phase & $\rho_1$--$\rho_2$
($\sigma^{-3}$) & $r_1$ ($\sigma$) & $z$ & parameters \\
\hline\hline
0--3.12 & fcc-I & 0--0.468 & 1.44574-- & 12 & --- \\
\hline
3.13--3.55 & bco & 0.549--0.559 & 1.07840--1.08786 & 2 & $1.326\div 1.330$,
$2.134\div 2.145$ \\
\hline 
3.56--4.62 & bct & 0.655--0.680 & 1.08276--1.09938 & 4 & $2.298\div 2.317$ \\
\hline 
4.63--5.54 & $\beta$Sn & 0.690--0.708 & 1.05191--1.06223 & 4 &
$0.729\div 0.733$ \\
\hline 
5.55--8.91 & sc & 0.801--0.859 & 1.05197--1.07677 & 6 & --- \\
\hline 
8.92--14.88 & sh & 0.940--1.017 & 1.03906--1.06239 & $\simeq 8$ &
$0.988\div 0.994$ \\
\hline
14.89-- & fcc-II & 1.166-- & --1.06645 & 12 & --- \\
\hline
\end{tabular*}
\end{table}

%
%
\begin{table}
\caption{\label{tab5}
Hertz potential: Zero-temperature phases up to $P=400$.
For $P>40$, only phases with 0 and 1 parameters were examined.
Notations are the same as in Table 2.
}
\begin{tabular*}{\columnwidth}[c]{@{\extracolsep{\fill}}|c|c|c|c|c|c|}
\hline
$P_1$--$P_2$ ($\epsilon\sigma^{-3}$) & phase & $\rho_1$--$\rho_2$
($\sigma^{-3}$) & $r_1$ ($\sigma$) & $z$ & parameters \\
\hline\hline
0--0.49 & fcc-I & 0--2.211 & 0.86161-- & 12 & --- \\
\hline
0.50--1.70 & bcc-I & 2.287--3.447 & 0.72232--0.82817 & 8 & --- \\
\hline 
1.71--2.45 & sh-I & 3.694--4.231 & 0.57655--0.60419 & 2 & $0.838\div 0.840$ \\
\hline 
2.46--3.96 & cI16-I & 4.353--5.249 & 0.51263--0.54563 & 3 & 0.125 \\
\hline 
3.97--5.73 & $\beta$Sn-I & 5.485--6.386 & 0.50503--0.52598 & 4 &
$2.416\div 2.508$ \\
\hline 
5.74--7.20 & bct-I & 6.456--7.116 & 0.45974--0.46571 & 2 & $0.571\div 0.588$ \\
\hline 
7.21--10.61 & A20-I & 7.366--8.651 & 0.48857--0.51514 & $\simeq 6$ &
$1.730\div 1.731$, $0.719\div 0.728$, 0.167 \\
\hline
10.62--11.10 & fluorite & 8.745--8.910 & 0.47819--0.48118 &
\parbox{0.05\textwidth}{\vspace{2mm}$8\times 1$, $4\times 2$\vspace{1mm}} &
--- \\
\hline
11.11--11.57 & fcc-II & 9.050--9.214 & 0.53541--0.53863 & 12 & --- \\
\hline
11.58--14.63 & A20-II & 9.331--10.309 & 0.49960--0.51378 & $\simeq 6$ & 1.731,
$1.448\div 1.475$, 0.167 \\
\hline
14.64--28.14 & bcc-II & 10.450--13.895 & 0.45386--0.49908 & 8 & --- \\
\hline
28.15--28.64 & $\beta$Sn-II & 14.093--14.202 & 0.40956--0.41061 & 4 &
0.578 \\
\hline
28.65--38.10 & sh-II & 14.254--16.214 & 0.38467--0.40155 & 6 &
$0.894\div 0.895$ \\
\hline
38.11--54.07 & sc & 16.417--19.265 & 0.37303--0.39346 & 6 & --- \\
\hline
54.08--58.89 & hR1-II & 19.431--20.218 & 0.37193--0.37660 & 6 &
$-0.074\div -0.072$ \\
\hline
58.90--75.02 & bct-II & 20.312--22.789 & 0.32347--0.33647 & 2 &
$0.617\div 0.622$ \\
\hline
75.03--102.78 & hR1-III & 22.951--26.643 & 0.34178--0.36507 & 6 &
$0.543\div 0.553$ \\
\hline
102.79--141.62 & hcp & 26.905--31.323 & 0.35608--0.37459 & 12 & --- \\
\hline
141.63--148.08 & bct-III & 31.413--32.096 & 0.34382--0.34639 & 8 &
$0.916\div 0.923$ \\
\hline
148.09--169.07 & bct-IV & 32.121--34.248 & 0.33691--0.34488 & 8 &
$1.116\div 1.155$ \\
\hline
169.08--206.32 & bct-V & 34.328--37.754 & 0.31071--0.32268 & 4 &
$1.734\div 1.766$ \\
\hline
206.33--242.37 & bcc12 & 37.822--40.873 & 0.31085--0.31900 & 8 & --- \\
\hline
242.38--307.75 & hR1-IV & 41.044--46.081 & 0.28614--0.30195 & 6 &
$0.130\div 0.171$ \\
\hline
307.76--345.22 & hR1-V & 46.113--48.746 & 0.27802--0.28422 & 6 &
$0.098\div 0.110$ \\
\hline
345.23--370.23 & $\beta$Sn-III & 48.859--50.541 & 0.27074--0.27429 & 4
& $3.967\div 3.988$ \\
\hline
370.24--400 & cI16-II & 50.669--52.602 & 0.25402--0.25721 & 3 & 0.078 \\
\hline
\end{tabular*}
\end{table}

\newpage
\begin{figure}
\includegraphics[width=16cm,angle=0]{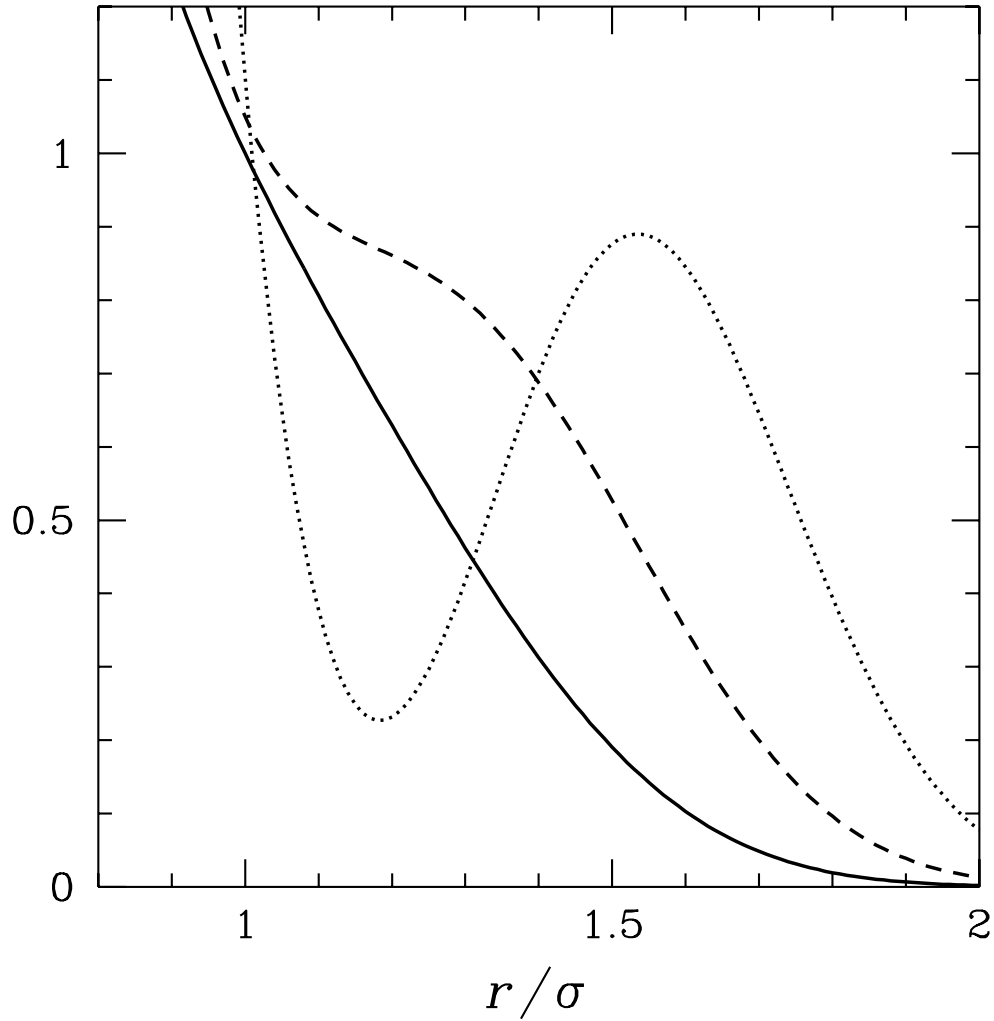}
\caption{\label{fig1}
Yoshida-Kamakura potential for $a=2.1$ (solid line, in units of $\epsilon$).
The figure reports also the force (dashed line, in units of $\epsilon/\sigma$)
and the second derivative of the potential (dotted line, in units of
$\epsilon/\sigma^2$). Note that the force and its steepness have been
divided by two and four, respectively, in order to fit into the picture.
}
\end{figure}

\newpage
\begin{figure}
\includegraphics[width=16cm,angle=0]{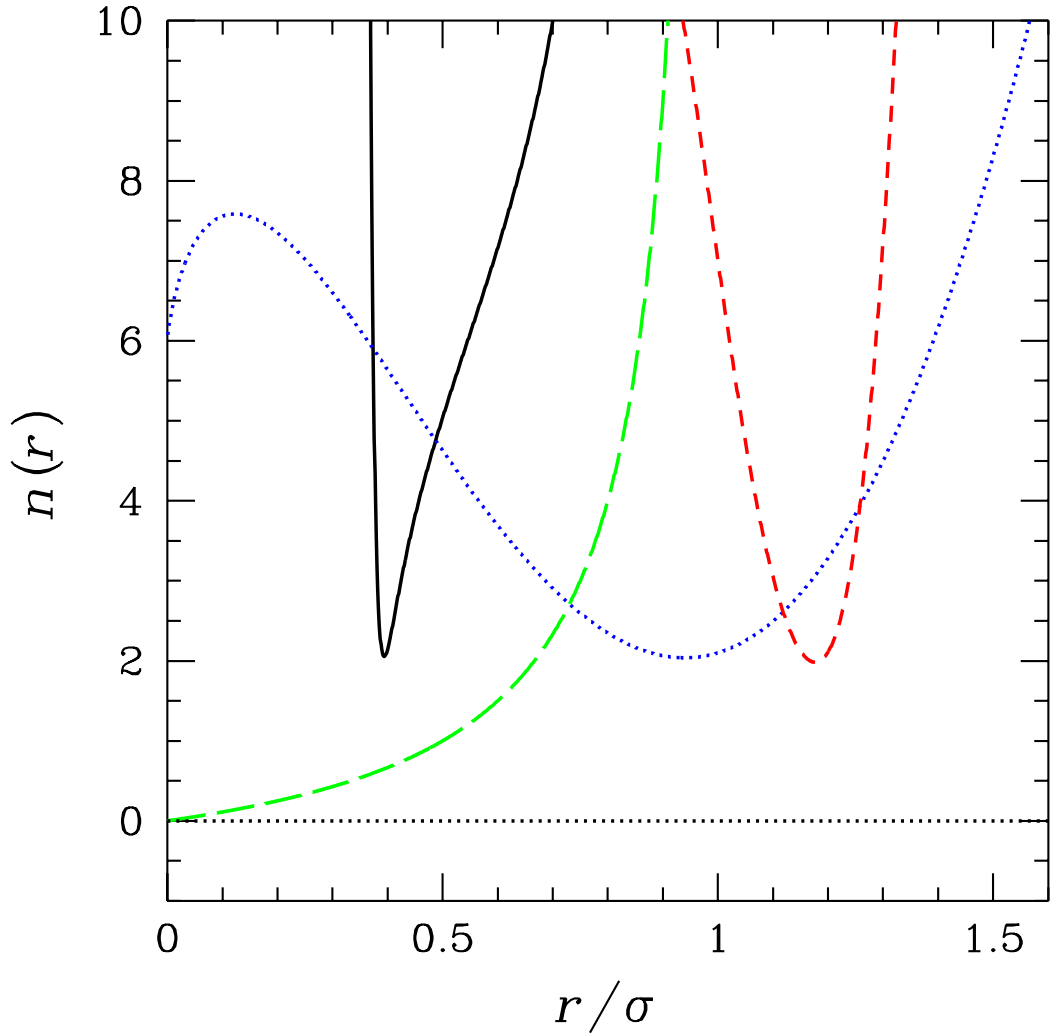}
\caption{\label{fig2}
(Color online) Effective inverse-power exponent $n(r)$ for the potentials
(\ref{2-1})-(\ref{2-4}), see Eq.\,(\ref{2-5}): Exp-6 potential (solid line),
Yoshida-Kamakura potential (dotted line), Fomin potential (dashed line),
and Hertz potential (long-dashed line). For the exp-6 case, $n(r)$ was
plotted only for those $r$ where $u_E(r)>0$.
}
\end{figure}

\newpage
\begin{figure}
\includegraphics[width=16cm,angle=0]{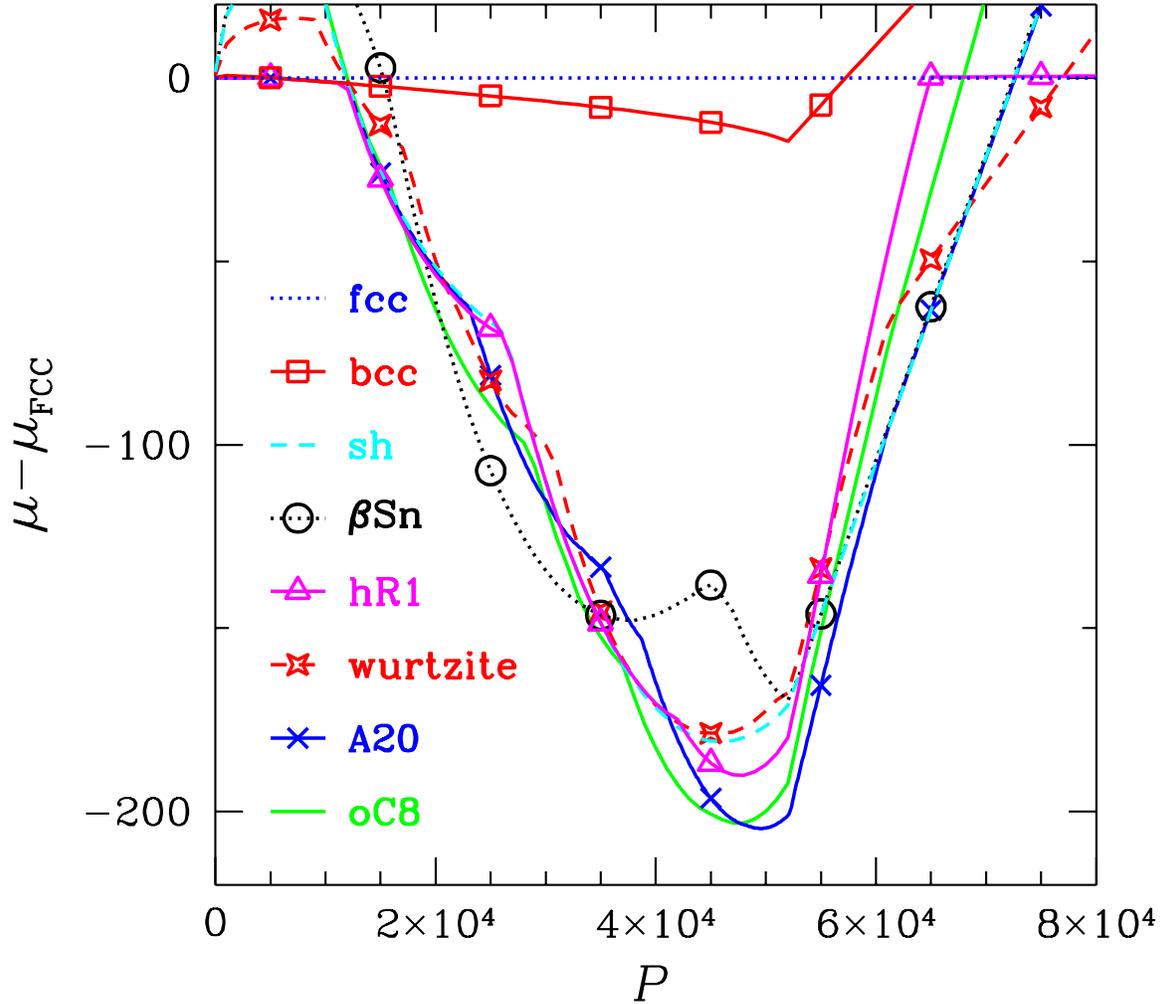}
\caption{\label{fig3}
(Color online) Exp-6 potential for $\alpha=11$: $T=0$ chemical potential
$\mu$, plotted as a function of pressure $P$, for a number of crystal
structures, choosing the fcc lattice as reference (both $\mu$ and $P$
are in reduced units; the chemical potentials of
structures that are never stable are not shown). Besided fcc, the stable
phases are bcc (squares and solid line), hR1 (triangles and solid line),
$\beta$Sn (dots and dotted line), oC8 (solid line), A20 (crosses and
solid line), sh (dashed line), and wurtzite (starred crosses and red
dashed line).
}
\end{figure}

\newpage
\begin{figure}
\includegraphics[width=16cm,angle=0]{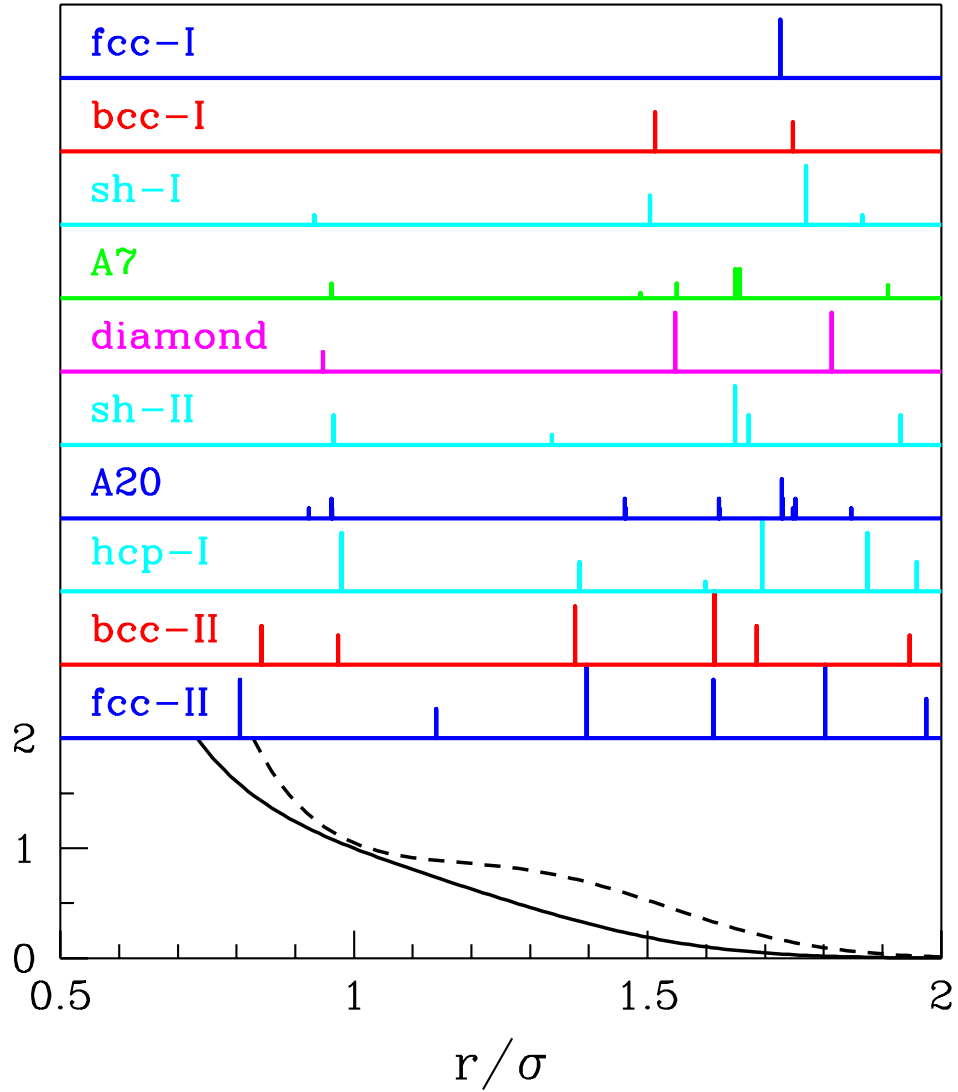}
\caption{\label{fig4}
(Color online) Yoshida-Kamakura potential: Comparison of the discrete
radial distribution function for the various zero-temperature phases
(the highest-pressure hcp-II and fcc-III phases are not shown).
At every distance, this function simply counts the number of neighbours at
that distance. The height of each panel is 15. Bottom panel: the potential
(solid line) and the corresponding force divided by two (dashed line).
From top to bottom: fcc-I ($P=0.32$), bcc-I ($P=0.95$),
sh-I ($P=1.79$), A7 ($P=2.43$), diamond ($P=3.73$),
sh-II ($P=5.19$), A20 ($P=8.77$), hcp-I ($P=13.88$),
bcc-II ($P=34$), and fcc-II ($P=56$).
}
\end{figure}

\newpage
\begin{figure}
\includegraphics[width=16cm,angle=0]{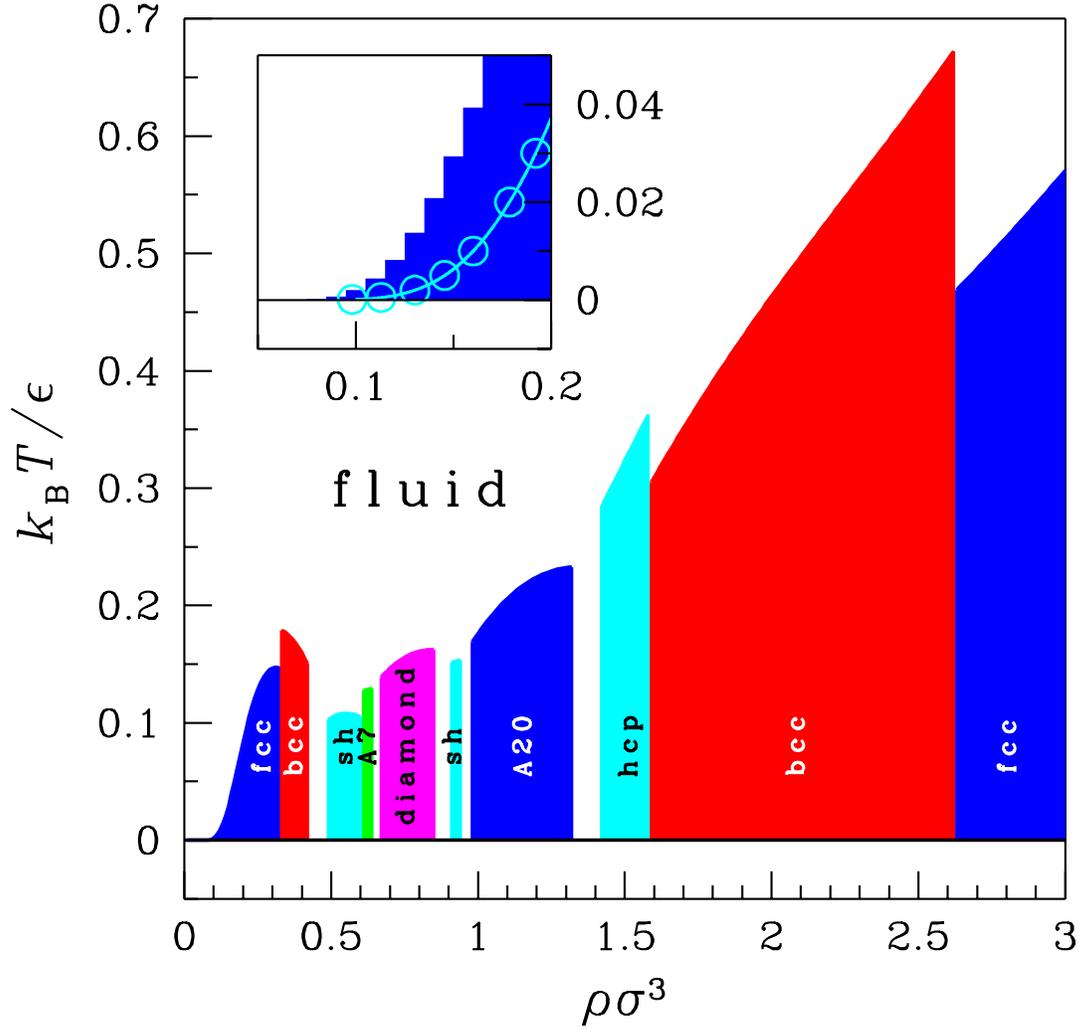}
\caption{\label{fig5}
(Color online) Yoshida-Kamakura potential: Schematic phase diagram in the
$\rho$-$T$ plane up to $\rho\sigma^3=3$, as obtained by combining the cell
theory for the $T=0$ solids with the Lindemann rule. Density gaps between
the phases are two-phase coexistence regions.
In the low-density region, the prediction of Eq.\,(\ref{2-10}) for the
fcc melting line (open dots) is compared with that based on the Lindemann
criterion.
}
\end{figure}
\end{document}